\documentclass[12pt,a4paper]{article}
\usepackage{amsmath}
\usepackage{latexsym}
\usepackage{amssymb}
\usepackage{graphicx,color}
\makeatletter
\def\rddots{\mathinner{\mkern1mu\raise\p@%
    \vbox{\kern7\p@\hbox{.}}\mkern2mu%
    \raise4\p@\hbox{.}\mkern2mu\raise7\p@\hbox{.}\mkern1mu}}
\makeatother
\makeatletter
\def\eqnarray{%
\stepcounter{equation}%
\let\@currentlabel=\theequation
\global\@eqnswtrue
\global\@eqcnt\z@
\tabskip\@centering
\let\\=\@eqncr
$$\halign to \displaywidth\bgroup\@eqnsel\hskip\@centering
$\displaystyle\tabskip\z@{##}$&\global\@eqcnt\@ne
\hfil$\displaystyle{{}##{}}$\hfil
&\global\@eqcnt\tw@$\displaystyle\tabskip\z@{##}$\hfil
\tabskip\@centering&\llap{##}\tabskip\z@\cr}
\makeatother
\newcommand{\ket}[1]{{\vert{#1}\rangle}}
\newcommand{\bra}[1]{{\langle{#1}\vert}}

\newcommand{\fukuso}{{\mathbf C}}

\begin{document}

\title{\sl General Solution of the Quantum Damped 
Harmonic Oscillator II : Some Examples}
\author{
  Kazuyuki FUJII
  \thanks{E-mail address : fujii@yokohama-cu.ac.jp }\quad and\ \ 
  Tatsuo SUZUKI
  \thanks{E-mail address : suzukita@aoni.waseda.jp ; 
  i027110@sic.shibaura-it.ac.jp }\\
  ${}^{*}$Department of Mathematical Sciences\\
  Yokohama City University\\
  Yokohama, 236--0027\\
  Japan\\
  ${}^{\dagger}$Center for Educational Assistance\\
  Shibaura Institute of Technology\\
  Saitama, 337--8570\\
  Japan\\
  }
\date{}
\maketitle
%
%
%
%
\begin{abstract}
  In the preceding paper (arXiv : 0710.2724 [quant-ph]) we have 
  constructed the general solution for the master equation of 
  quantum damped harmonic oscillator, which is given by the complicated 
  infinite series in the operator algebra level. 
  In this paper we give the explicit and compact forms to solutions 
  (density operators) for some initial values. 
  In particular, the compact one for the initial value based on 
  a coherent state is given, which has not been given as far as we know. 
  Moreover, some related problems are presented.
\end{abstract}
%


%
%
%
%

\vspace{5mm}
In the preceding paper \cite{EFS} (see also \cite{KF3}, \cite{KF4}) 
we treated the master equation of quantum damped harmonic oscillator
\begin{equation}
\label{eq:quantum damped harmonic oscillator}
\frac{\partial}{\partial t}\rho=-i[\omega a^{\dagger}a,\rho]
-
\frac{\mu}{2}
\left(a^{\dagger}a\rho+\rho a^{\dagger}a-2a\rho a^{\dagger}\right)
-
\frac{\nu}{2}
\left(aa^{\dagger}\rho+\rho aa^{\dagger}-2a^{\dagger}\rho{a}\right)
\end{equation}
where $\rho\equiv \rho(t)$ is the density operator (or matrix) of 
the system, and $a$ and $a^{\dagger}$ are the annihilation and creation 
operators of it (for example, an electro--magnetic field mode 
in a cavity), and $\mu$ and $\nu$ ($\mu > \nu \geq 0$) are some constants 
depending on it (for example, a damping rate of the cavity mode).

The general solution of the equation is given by
\begin{eqnarray}
\label{eq:general solution}
\rho(t)=
\frac{\mbox{e}^{\frac{\mu-\nu}{2}t}}{F(t)}
&&\sum_{n=0}^{\infty}
\frac{G(t)^{n}}{n!}(a^{\dagger})^{n}
\{
\exp\left(\{-i\omega t-\log(F(t))\}N\right)\times  \nonumber \\
&&
\left\{
\sum_{m=0}^{\infty}
\frac{E(t)^{m}}{m!}a^{m}\rho(0)(a^{\dagger})^{m}
\right\}
\exp\left(\{i\omega t-\log(F(t))\}N\right)
\}
a^{n}
\end{eqnarray}
where $N=a^{\dagger}a$ and 
\begin{eqnarray}
\label{eq:coefficients}
E(t)&=&\frac{\frac{2\mu}{\mu-\nu}\sinh\left(\frac{\mu-\nu}{2}t\right)}
     {\cosh\left(\frac{\mu-\nu}{2}t\right)+\frac{\mu+\nu}{\mu-\nu}
      \sinh\left(\frac{\mu-\nu}{2}t\right)},\quad
G(t)=\frac{\frac{2\nu}{\mu-\nu}\sinh\left(\frac{\mu-\nu}{2}t\right)}
     {\cosh\left(\frac{\mu-\nu}{2}t\right)+\frac{\mu+\nu}{\mu-\nu}
      \sinh\left(\frac{\mu-\nu}{2}t\right)}   \nonumber \\
F(t)&=&\cosh\left(\frac{\mu-\nu}{2}t\right)+
     \frac{\mu+\nu}{\mu-\nu}\sinh\left(\frac{\mu-\nu}{2}t\right).
\end{eqnarray}

If $\nu=0$, we have a simple form
\begin{equation}
\label{eq:corollary}
\rho(t)=
\mbox{e}^{-\left(\frac{\mu}{2}+i\omega\right)tN}
\left\{
\sum_{m=0}^{\infty}
\frac{\left(1-{e}^{-\mu t}\right)^{m}}{m!}a^{m}\rho(0)(a^{\dagger})^{m}
\right\}
\mbox{e}^{-\left(\frac{\mu}{2}-i\omega\right)tN}.
\end{equation}

\vspace{10mm}
Now we calculate $\rho(t)$ for some $\rho(0)$ given in the following.

\vspace{5mm}
\noindent
[A]\ \ $\rho(0)=\ket{0}\bra{0}$.\quad This case is easy and become
\begin{equation}
\rho(t)
=
\frac{\mbox{e}^{\frac{\mu-\nu}{2}t}}{F(t)}
\sum_{n=0}^{\infty}\frac{G(t)^{n}}{n!}(a^{\dagger})^{n}\ket{0}\bra{0}a^{n}
=
\frac{\mbox{e}^{\frac{\mu-\nu}{2}t}}{F(t)}
\sum_{n=0}^{\infty}G(t)^{n}\ket{n}\bra{n}
=
\frac{\mbox{e}^{\frac{\mu-\nu}{2}t}}{F(t)}e^{\log G(t) N}
\end{equation}
where $N\ (=a^{\dagger}a)$ is written as
\[
N=\sum_{n=0}^{\infty}n\ket{n}\bra{n}.
\]

\vspace{10mm}
\noindent
[B]\ \ $\rho(0)=\ket{\alpha}\bra{\alpha}$\ \ where  
$\ket{\alpha}$ ($\alpha \in \fukuso$) is a coherent state defined by 
\[
\ket{\alpha}
=e^{\alpha a^{\dagger}-\bar{\alpha}a}\ket{0}
=e^{-\frac{|\alpha|^{2}}{2}}e^{\alpha a^{\dagger}}\ket{0}.
\]
Note that
\[
a\ket{\alpha}=\alpha\ket{\alpha}\quad \mbox{and}\quad 
\bra{\alpha}a^{\dagger}=\bra{\alpha}\bar{\alpha}.
\]

\vspace{5mm}
\noindent
[B--1]\ \ Special Case : $\nu=0$.\quad From (\ref{eq:corollary}) we have
easily
\begin{equation}
\rho(t)
=
\ket{\alpha e^{-(\frac{\mu}{2}+i\omega)t}}
\bra{\alpha e^{-(\frac{\mu}{2}+i\omega)t}}.
\end{equation}
The proof is included in the next case. With respect to the evolution 
$\alpha\ \longrightarrow\ \alpha e^{-(\frac{\mu}{2}+i\omega)t}$, see 
the appendix and a review paper \cite{KH}.

\vspace{5mm}
\noindent
[B--2]\ \ General Case : $\nu\ne 0$.\quad We have

\begin{equation}
\label{eq:final form}
\rho(t)
=
\left(1-G(t)\right)
e^{|\alpha|^{2}e^{-(\mu-\nu)t}\log G(t)}
e^{-\log G(t)
\left\{
\alpha e^{-\frac{\mu-\nu}{2}t}e^{-i\omega t}a^{\dagger}+
\bar{\alpha}e^{-\frac{\mu-\nu}{2}t}e^{i\omega t}a-
N
\right\}
  }.
\end{equation}

Since this proof is not so easy we give a detailed one in the 
following.

\noindent
[First Step]\quad From (\ref{eq:general solution})
\[
\sum_{m=0}^{\infty}\frac{E(t)^{m}}{m!}a^{m}\ket{\alpha}\bra{\alpha}
(a^{\dagger})^{m}
=
\sum_{m=0}^{\infty}\frac{E(t)^{m}}{m!}{\alpha}^{m}\ket{\alpha}\bra{\alpha}
{\bar{\alpha}}^{m}
=
\sum_{m=0}^{\infty}\frac{(E(t)|\alpha|^{2})^{m}}{m!}\ket{\alpha}\bra{\alpha}
=
e^{E(t)|\alpha|^{2}}\ket{\alpha}\bra{\alpha}.
\]

\noindent
[Second Step]\quad From (\ref{eq:general solution}) we must calculate
\[
e^{\gamma N}\ket{\alpha}\bra{\alpha}e^{\bar{\gamma}N}
=
e^{\gamma N}e^{\alpha a^{\dagger}-\bar{\alpha}a}\ket{0}
\bra{0}e^{-(\alpha a^{\dagger}-\bar{\alpha}a)}e^{\bar{\gamma}N}
\]
where $\gamma=-i\omega t-\log(F(t))$. Note that $\bar{\gamma}\ne -\gamma$. 
It is easy to see
\[
e^{\gamma N}e^{\alpha a^{\dagger}-\bar{\alpha}a}\ket{0}
=
e^{\gamma N}e^{\alpha a^{\dagger}-\bar{\alpha}a}e^{-\gamma N}
e^{\gamma N}\ket{0}
=
e^{\alpha e^{\gamma}a^{\dagger}-\bar{\alpha}e^{-\gamma}a}\ket{0}
\]
where we used
\[
e^{\gamma N}a^{\dagger}e^{-\gamma N}=e^{\gamma}a^{\dagger}
\quad \mbox{and}\quad
e^{\gamma N}ae^{-\gamma N}=e^{-\gamma}a.
\]
See for example \cite{KF1}. Therefore by use of Baker--Campbell--Hausdorff 
formula two times
\begin{eqnarray*}
e^{\alpha e^{\gamma}a^{\dagger}-\bar{\alpha}e^{-\gamma}a}\ket{0}
&=&
e^{-\frac{|\alpha|^{2}}{2}}
e^{\alpha e^{\gamma}a^{\dagger}}e^{-\bar{\alpha}e^{-\gamma}a}\ket{0}
=
e^{-\frac{|\alpha|^{2}}{2}}e^{\alpha e^{\gamma}a^{\dagger}}\ket{0} \\
&=&
e^{-\frac{|\alpha|^{2}}{2}}e^{\frac{|\alpha|^{2}}{2}e^{\gamma +\bar{\gamma}}}
e^{\alpha e^{\gamma}a^{\dagger}-\bar{\alpha}e^{\bar{\gamma}}a}\ket{0}
=
e^{-\frac{|\alpha|^{2}}{2}(1-e^{\gamma +\bar{\gamma}})}\ket{\alpha e^{\gamma}}
\end{eqnarray*}
As a result we obtain
\[
e^{\gamma N}\ket{\alpha}\bra{\alpha}e^{\bar{\gamma}N}
=
e^{-|\alpha|^{2}(1-e^{\gamma +\bar{\gamma}})}
\ket{\alpha e^{\gamma}}\bra{\alpha e^{\gamma}}
\]
with $\gamma=-i\omega t-\log(F(t))$.

\noindent
[Third Step]\quad Under two steps above the equation 
(\ref{eq:general solution}) becomes 
\[
\rho(t)=
\frac{
e^{\frac{\mu-\nu}{2}t}
e^{|\alpha|^{2}(E(t)-1+e^{\gamma +\bar{\gamma}})}}{F(t)}
\sum_{n=0}^{\infty}
\frac{G(t)^{n}}{n!}(a^{\dagger})^{n}
\ket{\alpha e^{\gamma}}\bra{\alpha e^{\gamma}}a^{n}.
\]
We set $z=\alpha e^{\gamma}$ for simplicity and calculate 
\[
(\sharp)=\sum_{n=0}^{\infty}
\frac{G(t)^{n}}{n!}(a^{\dagger})^{n}\ket{z}\bra{z}a^{n}.
\]
Since $\ket{z}=e^{-|z|^{2}/2}e^{za^{\dagger}}\ket{0}$ we have
\begin{eqnarray*}
(\sharp)
&=&
e^{-|z|^{2}}
\sum_{n=0}^{\infty}
\frac{G(t)^{n}}{n!}(a^{\dagger})^{n}e^{za^{\dagger}}\ket{0}
\bra{0}e^{\bar{z}a}a^{n}
=
e^{-|z|^{2}}
e^{za^{\dagger}}
\left\{
\sum_{n=0}^{\infty}
\frac{G(t)^{n}}{n!}(a^{\dagger})^{n}\ket{0}\bra{0}a^{n}
\right\}
e^{\bar{z}a} \\
&=&
e^{-|z|^{2}}
e^{za^{\dagger}}
\left\{\sum_{n=0}^{\infty}G(t)^{n}\ket{n}\bra{n}\right\}
e^{\bar{z}a^{\dagger}}
=
e^{-|z|^{2}}e^{za^{\dagger}}e^{\log G(t)N}e^{\bar{z}a}
\end{eqnarray*}
by [A]. 
Namely, this form is a kind of disentangling formula, so 
we want to restore an entangling formula.

For that we use the {\bf disentangling formula}
\begin{equation}
\label{eq:formula-1}
e^{\alpha a^{\dagger}+\beta a+\gamma N}
=
e^{\alpha\beta\frac{e^{\gamma}-(1+\gamma)}{\gamma^{2}}}
e^{\alpha\frac{e^{\gamma}-1}{\gamma}a^{\dagger}}
e^{\gamma N}
e^{\beta\frac{e^{\gamma}-1}{\gamma}a}.
\end{equation}
For the proof see the fourth step in the following. From this 
it is easy to see
\begin{equation}
\label{eq:formula-2}
e^{xa^{\dagger}}e^{yN}e^{za}
=
e^{-\frac{xz(e^{y}-(1+y))}{(e^{y}-1)^{2}}}
e^{\frac{yx}{e^{y}-1}a^{\dagger}+\frac{yz}{e^{y}-1}a+yN}.
\end{equation}
Therefore
\[
(\sharp)=e^{-|z|^{2}}
e^{\frac{|z|^{2}(1+\log G(t)-G(t))}{(1-G(t))^{2}}}
e^{
\frac{\log G(t)}{G(t)-1}za^{\dagger}+
\frac{\log G(t)}{G(t)-1}\bar{z}a+
\log G(t)N
},
\]
so by noting
\[
z=\alpha e^{\gamma}=\alpha \frac{e^{-i\omega t}}{F(t)},\quad 
|z|^{2}=|\alpha|^{2}e^{\gamma + \bar{\gamma}},
\]
we have
\[
\rho(t)=
\frac{e^{\frac{\mu-\nu}{2}t}}{F(t)}
e^{|\alpha|^{2}(E(t)-1)}
e^{|\alpha|^{2}\frac{1+\log G(t)-G(t)}{F(t)^{2}(1-G(t))^{2}}}
e^{
\frac{\log G(t)}{F(t)(G(t)-1)}\alpha e^{-i\omega t}a^{\dagger}+
\frac{\log G(t)}{F(t)(G(t)-1)}\bar{\alpha} e^{i\omega t}a+
\log G(t)N
}.
\]

By the way, from (\ref{eq:coefficients}) 
\[
G(t)-1=-\frac{e^{\frac{\mu-\nu}{2}t}}{F(t)},\quad
\frac{1}{F(t)(G(t)-1)}=-e^{-\frac{\mu-\nu}{2}t},\quad
E(t)-1=-\frac{e^{-\frac{\mu-\nu}{2}t}}{F(t)}
\]
and
\begin{eqnarray*}
\frac{1-G(t)+\log G(t)}{F(t)^{2}(G(t)-1)^{2}}
&=&
e^{-(\mu-\nu)t}
\left\{
\frac{e^{\frac{\mu-\nu}{2}t}}{F(t)}+\log G(t)
\right\}
=
\frac{e^{-\frac{\mu-\nu}{2}t}}{F(t)}+e^{-(\mu-\nu)t}\log G(t) \\
&=&
-(E(t)-1)+e^{-(\mu-\nu)t}\log G(t)
\end{eqnarray*}
we finally obtain
\[
\rho(t)=
(1-G(t))e^{|\alpha|^{2}e^{-(\mu-\nu)t}\log G(t)}
e^{-\log G(t)
\left\{
\alpha e^{-i\omega t}e^{-\frac{\mu-\nu}{2}t}a^{\dagger}+
\bar{\alpha} e^{i\omega t}e^{-\frac{\mu-\nu}{2}t}a-
N
\right\}
}
\]
or 
\[
\rho(t)=
e^{|\alpha|^{2}e^{-(\mu-\nu)t}\log G(t)+\log (1-G(t))}
e^{-\log G(t)
\left\{
\alpha e^{-i\omega t}e^{-\frac{\mu-\nu}{2}t}a^{\dagger}+
\bar{\alpha} e^{i\omega t}e^{-\frac{\mu-\nu}{2}t}a-
N
\right\}
}.
\]

This is our main result in the paper. 

\noindent
[Fourth Step]\quad In last, let us give the proof to the disentangling 
formula (\ref{eq:formula-1}) because it is not so popular as far as 
we know.

Since
\begin{eqnarray*}
\alpha a^{\dagger}+\beta a+\gamma N
&=&
\gamma a^{\dagger}a+\alpha a^{\dagger}+\beta a \\
&=&
\gamma \left\{
\left(a^{\dagger}+\frac{\beta}{\gamma}\right)
\left(a+\frac{\alpha}{\gamma}\right)
-\frac{\alpha\beta}{\gamma^{2}}\right\}
=
\gamma 
\left(a^{\dagger}+\frac{\beta}{\gamma}\right)
\left(a+\frac{\alpha}{\gamma}\right)
-\frac{\alpha\beta}{\gamma}
\end{eqnarray*}
from (\ref{eq:formula-1}) we have
\begin{eqnarray*}
e^{\alpha a^{\dagger}+\beta a+\gamma N}
&=&
e^{-\frac{\alpha\beta}{\gamma}}
e^{\gamma 
\left(a^{\dagger}+\frac{\beta}{\gamma}\right)
\left(a+\frac{\alpha}{\gamma}\right)} \\
&=&
e^{-\frac{\alpha\beta}{\gamma}}
e^{\frac{\beta}{\gamma}a}
e^{\gamma a^{\dagger}
\left(a+\frac{\alpha}{\gamma}\right)} 
e^{-\frac{\beta}{\gamma}a}  \\
&=&
e^{-\frac{\alpha\beta}{\gamma}}
e^{\frac{\beta}{\gamma}a}
e^{-\frac{\alpha}{\gamma}a^{\dagger}}
e^{\gamma a^{\dagger}a}
e^{\frac{\alpha}{\gamma}a^{\dagger}}
e^{-\frac{\beta}{\gamma}a}.
\end{eqnarray*}
Then we obtain the disentangling formula (\ref{eq:formula-1}) 
($N=a^{\dagger}a$)
\begin{eqnarray*}
e^{-\frac{\alpha\beta}{\gamma}}
e^{\frac{\beta}{\gamma}a}
e^{-\frac{\alpha}{\gamma}a^{\dagger}}
e^{\gamma N}
e^{\frac{\alpha}{\gamma}a^{\dagger}}
e^{-\frac{\beta}{\gamma}a}
&=&
e^{-\frac{\alpha\beta}{\gamma}}
e^{-\frac{\alpha\beta}{\gamma^{2}}}
e^{-\frac{\alpha}{\gamma}a^{\dagger}}
e^{\frac{\beta}{\gamma}a}
e^{\gamma N}
e^{\frac{\alpha}{\gamma}a^{\dagger}}
e^{-\frac{\beta}{\gamma}a}  \\
&=&
e^{-(\frac{\alpha\beta}{\gamma}+
    \frac{\alpha\beta}{\gamma^{2}})}
e^{-\frac{\alpha}{\gamma}a^{\dagger}}
e^{\frac{\beta}{\gamma}a}
e^{\gamma N}
e^{\frac{\alpha}{\gamma}a^{\dagger}}
e^{-\frac{\beta}{\gamma}a}  \\
&=&
e^{-(\frac{\alpha\beta}{\gamma}+
    \frac{\alpha\beta}{\gamma^{2}})}
e^{-\frac{\alpha}{\gamma}a^{\dagger}}
e^{\gamma N}
e^{\frac{\beta}{\gamma}e^{\gamma}a}
e^{\frac{\alpha}{\gamma}a^{\dagger}}
e^{-\frac{\beta}{\gamma}a}  \\
&=&
e^{-(\frac{\alpha\beta}{\gamma}+
    \frac{\alpha\beta}{\gamma^{2}})
   +\frac{\alpha\beta}{\gamma^{2}}e^{\gamma}}
e^{-\frac{\alpha}{\gamma}a^{\dagger}}
e^{\gamma N}
e^{\frac{\alpha}{\gamma}a^{\dagger}}
e^{\frac{\beta}{\gamma}e^{\gamma}a}
e^{-\frac{\beta}{\gamma}a}  \\
&=&
e^{\alpha\beta\frac{e^{\gamma}-1-\gamma}{\gamma^{2}}}
e^{-\frac{\alpha}{\gamma}a^{\dagger}}
e^{\frac{\alpha}{\gamma}e^{\gamma}a^{\dagger}}
e^{\gamma N}
e^{\beta\frac{e^{\gamma}-1}{\gamma}a} \\
&=&
e^{\alpha\beta\frac{e^{\gamma}-1-\gamma}{\gamma^{2}}}
e^{\alpha\frac{e^{\gamma}-1}{\gamma}a^{\dagger}}
e^{\gamma N}
e^{\beta\frac{e^{\gamma}-1}{\gamma}a}
\end{eqnarray*}
by use of some commutation relations
\[
e^{sa}e^{ta^{\dagger}}=e^{st}e^{ta^{\dagger}}e^{sa},
\quad
e^{sa}e^{tN}=e^{tN}e^{se^{t}a},
\quad
e^{tN}e^{sa^{\dagger}}=e^{se^{t}a^{\dagger}}e^{tN},
\]
see for example \cite{KF1}.

\vspace{3mm}
We finished the proof. 
The formula (\ref{eq:final form}) that is compact and clear--cut 
has not been given as far as we know. See \cite{BP} and \cite{WS} 
as standard textbooks.

\vspace{3mm}
We are in a position to state our problem. A squeezed state 
$\ket{\beta}\ (\beta\in \fukuso)$ is defined as
\[
\ket{\beta}=e^{(1/2)\{\beta (a^{\dagger})^{2}-\bar{\beta}a^{2}\}}\ket{0}.
\]
See for example \cite{KF1}. For the initial value
\[
\rho(0)=\ket{\beta}\bra{\beta}
\]
we want to calculate $\rho(t)$ in (\ref{eq:general solution}) 
like in the text. However, we cannot sum up it in a compact 
form like (\ref{eq:final form}), so

\noindent
[{\bf Problem}]\ \ sum up $\rho(t)$ in a compact form.

Similarly, we can consider a coherent--squeezed state
\[
\ket{(\beta,\alpha)}=
e^{(1/2)\{\beta (a^{\dagger})^{2}-\bar{\beta}a^{2}\}}
e^{\alpha a^{\dagger}-\bar{\alpha}a}
\ket{0}
\]
for $(\beta,\alpha)\in \fukuso^{2}$ and treat the same problem 
for
\[
\rho(0)=\ket{(\beta,\alpha)}\bra{(\beta,\alpha)}.
\]

\noindent
They are important and interesting problems, and we leave them 
to readers.

\vspace{5mm}
In the paper \cite{EFS} we constructed the general solution of the 
quantum damped harmonic oscillator in the operator algebra level. 
It is given by some complicated infinite series, so it is desirable 
to sum up some solution with a special initial value in a compact form.

In this paper the compact form of the solution with the initial value 
based on coherent states was given. It is in fact fundamental, so it will 
be used in Quantum Open System or Quantum Optics in the near future.

However, we could not give a compact form to the solution with the 
initial value based on squeezed states or coherent--squeezed ones. 
We leave them to readers.

Lastly, we conclude the paper by stating our motivation. We are studying 
a model of quantum computation (computer) based on Cavity QED (see 
\cite{FHKW1} and \cite{FHKW2}), so in order to construct a more realistic 
model of (robust) quantum computer we have to study severe problems coming 
from decoherence. This is our future task.

\vspace{10mm}
\begin{center}
 \begin{Large}
  \textbf{Appendix}
 \end{Large}
\end{center}

\vspace{5mm}
In this appendix we review the solution of classical damped harmonic 
oscillator, which is important to understand the text. 
See any textbook on Mathematical Physics.

The differential equation is given by
\begin{equation}
\label{eq:classical damped harmonic oscillator}
\ddot{x}+\gamma \dot{x}+\omega^{2}x=0\quad (\gamma > 0)
\end{equation}
where $x=x(t),\ \dot{x}=dx/dt$ and the mass is set to 1 for simplicity. 
In the following we treat only the case $\omega > \gamma/2$ (the case 
$\omega=\gamma/2$ may be interesting). 

The solution is well--known to become
\begin{equation}
\label{eq:solution}
x(t)=e^{-\left(\frac{\gamma}{2}\pm i
\sqrt{\omega^{2}-(\frac{\gamma}{2})^{2}}\right)t}x(0)
\end{equation}
with complex form. If $\gamma/2\omega$ is small enough then we have
\begin{equation}
x(t)\approx 
e^{-\left(\frac{\gamma}{2}\pm i\omega\right)t}x(0)
=
x(0)e^{-\left(\frac{\gamma}{2}\pm i\omega\right)t}.
\end{equation}
%


\end{document}